\documentclass[conference]{IEEEtran}
\ifdefined\pdfoutput\pdfoutput=1\fi
\IEEEoverridecommandlockouts

\usepackage{iftex}
\ifPDFTeX
  \usepackage[T1]{fontenc}
  \usepackage{CJKutf8}
  \newcommand{\zh}[1]{\begin{CJK}{UTF8}{bsmi}#1\end{CJK}}
\else
  \usepackage{fontspec}
  \usepackage{xeCJK}
  \IfFontExistsTF{Noto Serif CJK TC}
    {\setCJKmainfont{Noto Serif CJK TC}}
    {\IfFontExistsTF{cwTeXQMing-Medium}
      {\setCJKmainfont{cwTeXQMing-Medium}}
      {\setCJKmainfont{FandolSong-Regular}}}
  \newcommand{\zh}[1]{#1}
\fi

\usepackage{amsmath,amssymb,bm}
\usepackage{booktabs,multirow,array}
\usepackage{graphicx}
\usepackage{tikz}
\usetikzlibrary{positioning,arrows.meta,fit,backgrounds,calc}
\usepackage{pgfplots}
\pgfplotsset{compat=1.17}
\usepackage{cite}
\usepackage{xurl}
\usepackage{microtype}
\usepackage{balance}
\usepackage{hyperref}

\tikzset{
  blk/.style={draw,rounded corners=2pt,align=center,inner sep=4pt,minimum height=8mm,font=\footnotesize},
  kept/.style={blk,fill=green!8,draw=green!55!black},
  swap/.style={blk,fill=blue!8,draw=blue!60!black},
  tok/.style={blk,fill=orange!10,draw=orange!70!black},
  io/.style={align=center,font=\footnotesize\itshape},
  ar/.style={-{Stealth[length=2mm]},semithick},
}

\makeatletter
\g@addto@macro\normalsize{%
  \setlength\abovedisplayskip{4pt plus 1pt minus 2pt}%
  \setlength\belowdisplayskip{4pt plus 1pt minus 2pt}%
  \setlength\abovedisplayshortskip{2pt plus 1pt}%
  \setlength\belowdisplayshortskip{2pt plus 1pt}%
}
\makeatother

\title{BlueMagpie-TTS: A Token-Efficient Tokenizer, Language Model, and TTS for Taiwanese-Accent Code-Switching Speech}

\author{\IEEEauthorblockN{Ho-Lam Chung, Bo-Xuan Zheng$^{\dagger}$, Cheng-Chieh Huang$^{\dagger}$, Cheng-Han Chang$^{\dagger}$,\\
Jung-Ching Chen$^{\dagger}$, Lok-Lam Ieong$^{\dagger}$, Ting-Lin Hsiao$^{\dagger}$, Yu-Cheng Lee$^{\dagger}$,\\
Yi-Hsin Chung$^{\dagger}$, Yu-Kai Guo$^{\dagger}$, Hung-yi Lee}
\IEEEauthorblockA{National Taiwan University}%
\thanks{$^{\dagger}$Equal contribution.}}

\begin{document}
\maketitle

\begin{abstract}
Off-the-shelf TTS systems are poorly adapted to Taiwanese Mandarin. Their accent defaults to other Mandarin variants, their tokenizers over-segment common Taiwanese text, and their pronunciation degrades at code-switching boundaries where Chinese and English alternate within one utterance. These problems share one root: the text side lacks adaptation to the Taiwanese context. We address the text side from the bottom up. PangolinTokenizer, a byte-level BPE tokenizer trained on Taiwan-context data, reaches the lowest token rate (0.485 tokens/character) with the smallest vocabulary among eight tokenizers. Barbet, a billion-parameter Traditional-Chinese language model trained on PangolinTokenizer, serves as the text-semantic frontend and ranks first among the compared billion-parameter models on a 14-task evaluation. BlueMagpie-TTS attaches Barbet to the pretrained acoustic stack of VoxCPM2 through a learned bridge, then jointly fine-tunes the assembled model on target-voice speech. On a 1{,}000-sentence Taiwan-localized test set, it lowers CER from 11.45\% to 4.81\%. It stays 25.2\% below a baseline that fine-tunes the same acoustic stack to the target voice, so the gain comes from the frontend, not from voice adaptation. In a blind listening study on 500 of these sentences with ten listeners, 65.6\% of majority votes prefer BlueMagpie-TTS.
\end{abstract}


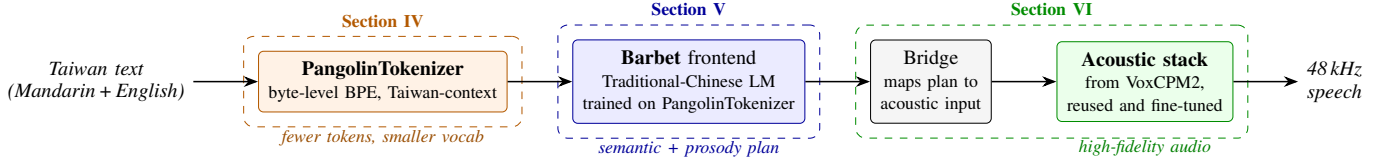
\begin{figure*}[t]
\centering
\resizebox{\textwidth}{!}{%
\begin{tikzpicture}[node distance=7mm and 9mm]
\node[io] (txt) {Taiwan text\\(Mandarin\,+\,English)};
\node[tok,right=of txt] (tok) {\textbf{PangolinTokenizer}\\\scriptsize byte-level BPE, Taiwan-context};
\node[swap,right=of tok] (lm) {\textbf{Barbet} frontend\\\scriptsize Traditional-Chinese LM\\\scriptsize trained on PangolinTokenizer};
\node[blk,right=of lm,fill=gray!8] (br) {Bridge\\\scriptsize maps plan to\\\scriptsize acoustic input};
\node[kept,right=of br] (ac) {\textbf{Acoustic stack}\\\scriptsize from VoxCPM2,\\\scriptsize reused and fine-tuned};
\node[io,right=of ac] (out) {48\,kHz\\speech};
\draw[ar] (txt)--(tok); \draw[ar] (tok)--(lm); \draw[ar] (lm)--(br); \draw[ar] (br)--(ac); \draw[ar] (ac)--(out);
\node[font=\scriptsize\itshape,orange!60!black,below=1.2mm of tok] {fewer tokens, smaller vocab};
\node[font=\scriptsize\itshape,blue!60!black,below=1.2mm of lm] {semantic + prosody plan};
\node[font=\scriptsize\itshape,green!50!black,below=1.2mm of ac] {high-fidelity audio};
\begin{scope}[on background layer]
\node[draw=orange!70!black,dashed,rounded corners,fit=(tok),inner sep=2mm,label={[font=\scriptsize\bfseries,orange!70!black]above:Section~\ref{sec:tok}}] {};
\node[draw=blue!60!black,dashed,rounded corners,fit=(lm),inner sep=2mm,label={[font=\scriptsize\bfseries,blue!60!black]above:Section~\ref{sec:lm}}] {};
\node[draw=green!55!black,dashed,rounded corners,fit=(br)(ac),inner sep=2mm,label={[font=\scriptsize\bfseries,green!55!black]above:Section~\ref{sec:tts}}] {};
\end{scope}
\end{tikzpicture}%
}
\caption{The Taiwan-localized stack, built around token efficiency. PangolinTokenizer (orange) represents Taiwan text at a low token rate. Barbet (blue), a Traditional-Chinese language model trained on PangolinTokenizer, plans what to say and how to say it. BlueMagpie-TTS (green) attaches Barbet to the reused VoxCPM2 acoustic stack through a bridge module and fine-tunes the assembled model.}
\label{fig:stack}
\end{figure*}

\section{Introduction}
\label{sec:intro}
Neural TTS now produces near-human speech for major languages on standard reading benchmarks~\cite{vits,seedtts,f5tts,voxcpm}. Most of these systems, however, are poorly adapted to Taiwanese Mandarin. Their accent defaults to other regional Mandarin variants, and Taiwanese listeners perceive the mismatch immediately. Their text side is not trained on the terms and writing conventions common in Taiwan, so the model has limited capacity for local usage. These two gaps, the accent and the text representation, reduce the naturalness of off-the-shelf TTS for Taiwanese Mandarin speakers.

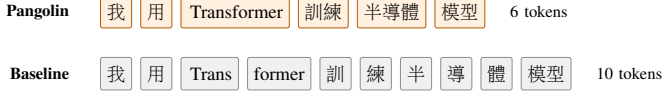
\begin{figure}[t]
\centering
\resizebox{\columnwidth}{!}{%
\begin{tikzpicture}[
  tk/.style={draw,rounded corners=1pt,minimum height=5.5mm,inner xsep=3pt,font=\small,fill=orange!12,draw=orange!70!black},
  tb/.style={draw,rounded corners=1pt,minimum height=5.5mm,inner xsep=3pt,font=\small,fill=black!6,draw=black!45},
  lab/.style={font=\footnotesize\bfseries,anchor=east},
  node distance=1.4mm,
]
\node[lab] (lp) at (0,0) {Pangolin};
\node[tk,right=4mm of lp] (p1) {\zh{我}};
\node[tk,right=of p1] (p2) {\zh{用}};
\node[tk,right=of p2] (p3) {\,Transformer};
\node[tk,right=of p3] (p4) {\zh{訓練}};
\node[tk,right=of p4] (p5) {\zh{半導體}};
\node[tk,right=of p5] (p6) {\zh{模型}};
\node[right=3mm of p6,font=\footnotesize] {6 tokens};
\node[lab] (lb) at (0,-1.05) {Baseline};
\node[tb,right=4mm of lb] (b1) {\zh{我}};
\node[tb,right=of b1] (b2) {\zh{用}};
\node[tb,right=of b2] (b3) {\,Trans};
\node[tb,right=of b3] (b4) {former};
\node[tb,right=of b4] (b5) {\zh{訓}};
\node[tb,right=of b5] (b6) {\zh{練}};
\node[tb,right=of b6] (b7) {\zh{半}};
\node[tb,right=of b7] (b8) {\zh{導}};
\node[tb,right=of b8] (b9) {\zh{體}};
\node[tb,right=of b9] (b10) {\zh{模型}};
\node[right=3mm of b10,font=\footnotesize] {10 tokens};
\end{tikzpicture}%
}
\caption{Illustrative tokenization of one code-switching sentence. A Taiwan-context tokenizer keeps Taiwan terms (\zh{半導體}) and the English word as fewer pieces, so the token sequence is shorter than a general multilingual tokenizer on the same text. Aggregate rates over PangolinBench are in Table~\ref{tab:compression}.}
\label{fig:toks}
\end{figure}

The text side is where the problem starts. Most TTS systems build on a general multilingual tokenizer~\cite{sennrich2016bpe}. Such a tokenizer represents Taiwan-context text losslessly, but it often fragments common terms into many tokens~\cite{rust2021tokenizer}. A term such as \zh{半導體} costs one token under a Taiwan-context tokenizer, but several under a general multilingual one. Figure~\ref{fig:toks} shows the effect on one sentence. The fragmentation wastes sequence length, raises inference cost, and means the language model trains on a longer, noisier representation of the same content.

Code-switching makes it worse. A single spoken sentence in Taiwan often mixes Mandarin with English words, abbreviations, and proper nouns. A speaker says \zh{我用} Transformer \zh{訓練模型}, one Mandarin sentence with an English word in the middle. At the switch point, the tokenizer fragments the boundary, the language model cannot plan the prosody transition, and the TTS distorts the English span or breaks the surrounding Mandarin prosody. Chinese--English mixing is common in education, technology, and media, so these failures are not rare.

Fixing these problems requires adapting the text side from the bottom up. The first step is a tokenizer trained on Taiwan-context text, so that common terms stay as single tokens and the sequence is short. The second step is a language model trained on this tokenizer, so that the model understands Taiwanese usage and can plan prosody across code-switching boundaries. The third step is connecting this frontend to a TTS acoustic stack, so that the accent and the switch points are rendered correctly. We build one stack following this recipe, shown in Figure~\ref{fig:stack}. Section~\ref{sec:overview} gives the overview.

Our contributions follow the same bottom-up order.
\begin{itemize}
\itemsep0.15em
\item We build \textbf{PangolinTokenizer}, a byte-level BPE tokenizer for Taiwan-context text and speech transcripts, with \textbf{PangolinBench}, a tokenizer benchmark with ten Taiwan-focused subsets (Section~\ref{sec:tok}). It reaches 0.485 tokens/character, the lowest among eight tokenizers, with the smallest vocabulary.
\item We train \textbf{Barbet}, a billion-parameter Traditional-Chinese language model on PangolinTokenizer, and use it as the text-semantic frontend (Section~\ref{sec:lm}). It ranks first among the compared billion-parameter public models on the TAIDE-14 task set~\cite{taide14} at 0.7488 bits per byte.
\item We present \textbf{BlueMagpie-TTS}, a Taiwanese-accent code-switching TTS built by attaching Barbet to a reused acoustic stack through a learned bridge (Section~\ref{sec:tts}). On a 1{,}000-sentence Taiwan-localized test set it lowers CER from 11.45\% to 4.81\%, and listeners prefer it on 65.6\% of sentences (Section~\ref{sec:exp}).
\end{itemize}

\section{Related Work}
\label{sec:related}
\noindent\textbf{Tokenization.}
Subword tokenization is standard for neural sequence models. BPE merges frequent symbol pairs~\cite{sennrich2016bpe}, the unigram model offers a different objective~\cite{kudo2018subword}, and SentencePiece trains either scheme from raw text~\cite{kudo2018sentencepiece}. Byte-level BPE operates on UTF-8 bytes and guarantees a lossless round-trip~\cite{radford2019gpt2}. Most large models use a large multilingual vocabulary~\cite{grattafiori2024llama3,yang2024qwen25,gemmateam2025gemma3}. A tokenizer trained for one language can beat a multilingual tokenizer on that language~\cite{rust2021tokenizer}, and token cost differs across languages, which raises a fairness concern~\cite{petrov2023unfairness,ahia2023cost}. Other work studies intrinsic tokenizer metrics that predict downstream quality~\cite{zouhar2023noiseless}. PangolinBench follows this intrinsic direction, with explicit pass criteria for Taiwan-context text. PangolinTokenizer follows the byte-level BPE design for Taiwan-context text. Token-free models instead read raw bytes or characters~\cite{xue2022byt5,clark2022canine}. They keep a lossless round-trip but produce longer sequences. We keep a learned vocabulary for a lower token count, and keep the byte-level base for the lossless guarantee.

\noindent\textbf{Decoupled semantic--acoustic TTS.}
Neural TTS has moved through several designs. Early systems paired an acoustic model with a vocoder~\cite{tacotron2,fastspeech2}, and end-to-end models merged the two stages~\cite{vits}. Recent systems cast speech generation as language modeling. Neural codec language models predict audio tokens from text~\cite{valle}, and diffusion or flow-matching decoders render audio~\cite{f5tts,seedtts}. VoxCPM removes the external tokenizer and learns a continuous representation. It uses a text-semantic language model that plans semantics and prosody, and a residual acoustic model that renders audio~\cite{voxcpm,voxcpm2}. This split puts language understanding and prosody planning in the text-semantic model, and audio rendering in the acoustic model. We treat the text-semantic model as a replaceable frontend.

\noindent\textbf{Code-switching and Taiwan-localized models.}
Multilingual zero-shot systems cover many languages from one model~\cite{xtts,seedtts}. Prior work builds Mandarin--English code-switching TTS with cross-lingual language models~\cite{cstts}. That work targets the switching, not a Taiwanese accent, and it retrains acoustic models. Self-supervised models learn speech representations from unlabeled audio~\cite{wav2vec2,hsu2021hubert}, which underpin many speech systems. Taiwan-focused efforts release Traditional-Chinese language models~\cite{taiwanllm} and understanding benchmarks~\cite{tmlu}, and ASR for Taiwanese Mandarin and code-switching~\cite{breezeasr}. These resources show that Taiwan-context language and speech benefit from targeted data and models, which motivates our Taiwan-context tokenizer and frontend. We use such an ASR, built on a large multilingual backbone~\cite{radford2023whisper}, as an evaluation tool.

\section{Overview}
\label{sec:overview}
Figure~\ref{fig:stack} shows the design. The system has three parts: PangolinTokenizer, Barbet, and a reused acoustic stack from VoxCPM2~\cite{voxcpm2}. Each part is trained or reused in a fixed order, and the order determines what each part can rely on.

\textbf{PangolinTokenizer} defines the vocabulary. Barbet is trained on this vocabulary, so the token granularity of PangolinTokenizer propagates to every downstream component. A shorter token sequence means a shorter plan for the frontend and a lower training and inference cost for the assembled model. PangolinTokenizer is trained once and then frozen.

\textbf{Barbet} is a billion-parameter Traditional-Chinese language model trained on PangolinTokenizer. It reads the input text and produces a semantic-prosodic plan: what to say and how to say it. The accent and the code-switching decisions live in this plan, so localizing the frontend is what localizes the speech. Barbet is pretrained on text, then adapted to the speech task during joint fine-tuning.

\textbf{VoxCPM2 acoustic stack} renders the plan into a 48~kHz waveform. It already produces high-fidelity audio, so we reuse its pretrained weights rather than training an acoustic model from scratch. A bridge module maps Barbet's hidden states into the input space of the acoustic stack, because the two pretrained models use different representations. We first train the bridge to align the two spaces, then jointly fine-tune the assembled model, Barbet plus bridge plus acoustic stack, on target-voice speech. Joint fine-tuning lets the frontend and the acoustic stack co-adapt: Barbet learns to produce plans that the acoustic stack can render, and the acoustic stack learns to follow Barbet's plan for Taiwan-accent prosody.

Section~\ref{sec:tok} describes PangolinTokenizer, Section~\ref{sec:lm} describes Barbet, and Section~\ref{sec:tts} describes the bridge and the joint fine-tuning.

\section{PangolinTokenizer}
\label{sec:tok}
Barbet is trained on PangolinTokenizer. We now describe the tokenizer and why its token efficiency matters for the frontend and for the wider Taiwan-context text setting.

\subsection{Why a Taiwan-specific tokenizer}
Chinese writing has no whitespace word boundaries, so the merge rules alone decide how many tokens a common word costs. If the merge rules under-cover Traditional Chinese, common words break into short fragments, which raises token count and lowers usable context. Taiwanese text also contains many local terms, such as \zh{健保}, \zh{戶政事務所}, \zh{捷運}, and \zh{晶圓代工}. If the merge rules do not learn them, the model pays a higher token cost for the same content. As Figure~\ref{fig:toks} shows, the effect is consistent across Taiwan-context text.

Taiwanese text also mixes writing systems. Real text contains Traditional Chinese, Bopomofo, Taigi Han characters and romanization (for example T\^{a}i-g\'{i}, tshit-\'{a}), Hakka romanization, emoji, URLs, and JSON. Romanized Taigi and Hakka are not plain English. They use accented Latin letters and hyphen-separated syllables. A tokenizer tuned for English can segment them unstably, which raises the token count on these spans.

\subsection{Byte-level BPE for lossless round-trip}
PangolinTokenizer uses the 256 byte values as the base representation, then learns BPE merges for frequent fragments~\cite{sennrich2016bpe,radford2019gpt2}. The tokenizer does not enumerate Unicode in advance. It also avoids the \texttt{<unk>} problem of character-level tokenizers on unseen characters. Lossless round-trip means that encoding then decoding returns the input exactly, with no lost characters. On the curated PangolinBench samples, which span Traditional Chinese, Bopomofo, romanized Taigi and Hakka, emoji, JSON, URLs, and speech transcripts, round-trip accuracy is 100\%.

\subsection{Taiwan-context merge rules}
A tokenizer vocabulary reflects its training distribution. A tokenizer trained mostly on general multilingual data can sacrifice token efficiency on a specific local context~\cite{rust2021tokenizer}. PangolinTokenizer does not aim for the best average across all languages. It aims for a low token cost on Taiwan-context text. The training corpus is stratified by language, script, domain, and format. It covers formal Traditional Chinese, Taiwanese Mandarin speech, named entities, Taigi, Hakka, Bopomofo, code-switching technical text, and ASR transcripts. The merge rules therefore spend their budget on the fragments that Taiwan-context text uses most, not on a global average.

\subsection{Training pipeline}
Taiwan-context corpora mix Traditional Chinese, Bopomofo, romanized Taigi and Hakka, code-switching, and structured transcripts. The corpus is large, about 20 billion tokens, and the mixture is heterogeneous, so loading everything into memory before training is impractical. We implement a custom streaming byte-level BPE trainer that learns merge rules from corpus pair frequencies in a single pass, without holding the full corpus in memory. It splits large files into line-aligned chunks, counts fragment frequencies in parallel workers, and merges the statistics in the main process, so memory stays bounded as the corpus grows. The trainer is deterministic: the same corpus and parameters produce the same merge order, hence the same tokenizer, which supports reproducibility and auditing. The output is a Hugging Face compatible tokenizer, loadable without custom code. The streaming design also lets us iterate on the corpus mixture quickly, because re-running with a different mix does not require more memory, only more time.

\subsection{PangolinBench and results}
\begin{figure}[t]
  \centering
  \includegraphics[width=0.92\linewidth]{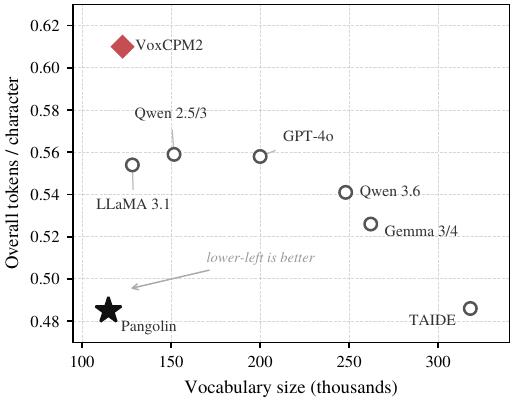}
  \caption{Vocabulary size against overall tokens/character. Lower-left is better. PangolinTokenizer (star) reaches the lowest token rate with the smallest vocabulary. VoxCPM2 (diamond) has a similar vocabulary but a 26\% higher token rate.}
  \label{fig:pareto}
\end{figure}

PangolinTokenizer reaches the lowest overall token rate among eight tokenizers, 0.485 tokens/character, with the smallest vocabulary (Table~\ref{tab:compression}). Against the replaced VoxCPM2 tokenizer, it lowers the overall rate by 20\% (0.485 vs.\ 0.610) and cuts the Taiwan lexicon weighted cost by 42\% (Table~\ref{tab:lexicon}). On a vocabulary efficiency score that separates compression from vocabulary size (Eq.~\ref{eq:ve}), PangolinTokenizer reaches 1.795, the highest of all, 1.27 times the second-best (Table~\ref{tab:efficiency}). Figure~\ref{fig:pareto} shows the Pareto view: PangolinTokenizer is the only non-dominated point, with the smallest vocabulary and the lowest token rate at once.

We measure these results on PangolinBench, a tokenizer benchmark that evaluates whether a tokenizer is a suitable text foundation for a Taiwan-localized TTS. In a TTS system, the tokenizer determines how efficiently the frontend represents the input, how well it preserves Taiwan-specific terms, and how cleanly it handles code-switching boundaries. PangolinBench measures these through ten subsets covering formal Traditional Chinese, Taiwanese Mandarin, Taigi, Hakka, Bopomofo, code-switching, and structured text. Each metric group has explicit pass criteria. We compare PangolinTokenizer against seven baseline tokenizers: six from widely used language models~\cite{grattafiori2024llama3,yang2024qwen25,gemmateam2025gemma3}, and the VoxCPM2 tokenizer that BlueMagpie-TTS replaces. Among the baselines, TAIDE is the tokenizer of Taiwan's national language-model project~\cite{taide}, which extends a base vocabulary with Traditional-Chinese entries, so it is the strongest baseline on Taiwan text.

On the core Taiwanese Mandarin subsets, PangolinTokenizer is second only to TAIDE, while using 36\% of its vocabulary. On formal Traditional Chinese it lowers the token count by 25\% against GPT-4o and LLaMA~3.1; on Taiwan named entities, by about 28\% against GPT-4o (0.730 vs.\ 1.009). These are the text types Taiwan-context applications use most, so the saving compounds over a corpus. PangolinTokenizer also reaches the lowest rate on structured text, 0.400 on rich transcription JSON. The VoxCPM2 tokenizer scores 0.922 on formal Traditional Chinese, the worst among all tokenizers, and 0.982 on Taiwan named entities, among the worst, because its character-level Chinese design fragments every multi-character term. Since Taiwanese Mandarin speech is mostly Chinese text, this fragmentation directly raises the frontend cost in the TTS pipeline. The VoxCPM2 tokenizer does better on code-switching (0.457 vs.\ 0.478), because its English vocabulary is not character-level, but the net effect is a 26\% higher overall token rate. On the harder subsets, a 114{,}822 vocabulary cannot allocate enough merges to all scripts at once, so larger-vocabulary tokenizers reach lower rates, and a larger vocabulary is one way to close the remaining gap. Round-trip accuracy stays at 100\% on every subset, so a lower token count never costs a lost character.

A tokenizer can reach a low rate by using a large vocabulary, so we separate the two effects with a vocabulary efficiency score,
\begin{equation}
  \mathrm{Efficiency} = \frac{1 \,/\, (\text{tokens per character})}{(\text{vocabulary size}) \,/\, 100000} .
  \label{eq:ve}
\end{equation}
The score is an empirical formula that puts token rate and vocabulary size on one scale, not a derived quantity. Every other tokenizer pays both a larger vocabulary and a higher token rate. TAIDE reaches a close overall rate (0.486) but with a vocabulary 2.77 times larger, which costs more embedding and output-projection parameters. A smaller vocabulary also reduces these parameters, and a lower token rate shortens sequences at inference.

\begin{table*}[t]
  \caption{Token compression on PangolinBench, reported as tokens/character unless noted. Lower is better. \textbf{Best} in bold, \underline{second-best} underlined. VoxCPM2 is the tokenizer replaced by PangolinTokenizer in BlueMagpie-TTS. Several models share one tokenizer: Qwen2.5/Qwen3 and Gemma3/Gemma4.}
  \label{tab:compression}
  \centering
  \setlength{\tabcolsep}{3.4pt}
  \resizebox{\textwidth}{!}{%
  \begin{tabular}{l cccccccc}
    \toprule
    \textbf{Subset} & \textbf{Pangolin} & \textbf{VoxCPM2} & \textbf{TAIDE} & \textbf{Gemma 3/4} & \textbf{Qwen 3.6} & \textbf{LLaMA 3.1} & \textbf{GPT-4o} & \textbf{Qwen 2.5/3} \\
    \midrule
    Vocabulary size            & 114{,}822 & 122{,}753 & 318{,}080 & 262{,}144 & 248{,}070 & 128{,}256 & 200{,}019 & 151{,}669 \\
    \midrule
    Formal Traditional Chinese & \underline{0.676} & 0.922 & \textbf{0.559} & 0.716 & \underline{0.676} & 0.902 & 0.902 & 0.735 \\
    Taiwanese Mandarin colloquial & \underline{0.708} & 0.979 & \textbf{0.688} & 0.750 & 0.750 & 0.917 & 0.979 & 0.792 \\
    Taiwan named entities      & \underline{0.730} & 0.982 & \textbf{0.622} & 0.892 & 0.757 & 1.000 & 1.009 & 0.919 \\
    Taigi Han-roman mixed      & 0.714 & 0.703 & \textbf{0.538} & 0.582 & \underline{0.615} & 0.637 & \underline{0.615} & 0.692 \\
    Hakka romanized (tokens/byte) & 0.372 & 0.361 & \textbf{0.293} & \underline{0.314} & 0.340 & 0.335 & 0.319 & 0.346 \\
    Bopomofo                   & 1.068 & 1.364 & \textbf{0.750} & \underline{0.795} & 1.386 & 1.205 & 1.227 & 1.114 \\
    Code-switching             & 0.478 & 0.457 & \textbf{0.380} & \textbf{0.380} & \underline{0.413} & 0.424 & 0.457 & 0.457 \\
    Rich transcription JSON    & \textbf{0.400} & 0.468 & 0.423 & 0.428 & 0.441 & \underline{0.405} & \underline{0.405} & 0.447 \\
    \midrule
    \textbf{Overall tokens/character} & \textbf{0.485} & 0.610 & \underline{0.486} & 0.526 & 0.541 & 0.554 & 0.558 & 0.559 \\
    \bottomrule
  \end{tabular}%
  }
\end{table*}

\begin{table}[t]
  \caption{Vocabulary efficiency, defined in Eq.~\ref{eq:ve}. Higher is better. \textbf{Best} in bold, \underline{second-best} underlined.}
  \label{tab:efficiency}
  \centering
  \setlength{\tabcolsep}{5pt}
  \begin{tabular}{l r r r}
    \toprule
    \textbf{Tokenizer} & \textbf{Vocab.} & \textbf{tok/char} & \textbf{Efficiency} \\
    \midrule
    PangolinTokenizer & 114{,}822 & 0.485 & \textbf{1.795} \\
    LLaMA 3.1         & 128{,}256 & 0.554 & \underline{1.408} \\
    Qwen 2.5/3        & 151{,}669 & 0.559 & 1.179 \\
    GPT-4o            & 200{,}019 & 0.558 & 0.896 \\
    Qwen 3.6          & 248{,}070 & 0.541 & 0.745 \\
    Gemma 3/4         & 262{,}144 & 0.526 & 0.725 \\
    TAIDE             & 318{,}080 & 0.486 & 0.647 \\
    VoxCPM2           & 122{,}753 & 0.610 & 1.335 \\
    \bottomrule
  \end{tabular}
\end{table}

\subsection{Taiwan lexicon cost}
PangolinBench measures a weighted token cost and an over-fragmentation rate on a set of Taiwan terms. Frequent terms such as \zh{健保}, \zh{捷運}, \zh{台北}, \zh{臺南}, and \zh{半導體} encode to a single token under PangolinTokenizer. Longer compounds still split, for example \zh{晶圓代工} into four tokens, which leaves room for further merge-rule tuning. Table~\ref{tab:lexicon} reports the comparison. TAIDE reaches the lowest weighted cost, in line with its 318K vocabulary. PangolinTokenizer is second at 2.48. Against LLaMA~3.1, which has a similar vocabulary size, it lowers the weighted Taiwan lexicon cost by about 29\% (2.48 vs.\ 3.47). Under a limited vocabulary, Taiwan-context merge rules reduce fragmentation of Taiwan terms. The over-fragmentation rate tells the same story. PangolinTokenizer splits 27\% of Taiwan terms into more than three tokens, against 40\% for GPT-4o and 53\% for LLaMA~3.1, so fewer terms are broken into many pieces.

\begin{table}[t]
  \caption{Taiwan lexicon weighted token cost and over-fragmentation rate (fraction of terms split into more than three tokens). Lower is better for both. \textbf{Best} in bold, \underline{second-best} underlined.}
  \label{tab:lexicon}
  \centering
  \small
  \setlength{\tabcolsep}{4pt}
  \begin{tabular}{l r r r}
    \toprule
    \textbf{Tokenizer} & \textbf{Vocab.} & \textbf{Weighted cost} & \textbf{Over-frag.} \\
    \midrule
    TAIDE             & 318{,}080 & \textbf{1.70} & \textbf{20\%} \\
    PangolinTokenizer & 114{,}822 & \underline{2.48} & \underline{27\%} \\
    Qwen 3.6          & 248{,}070 & 2.59 & \underline{27\%} \\
    Gemma 3/4         & 262{,}144 & 2.92 & 33\% \\
    Qwen 2.5/3        & 151{,}669 & 3.08 & 47\% \\
    GPT-4o            & 200{,}019 & 3.45 & 40\% \\
    LLaMA 3.1         & 128{,}256 & 3.47 & 53\% \\
    VoxCPM2           & 122{,}753 & 4.31 & 60\% \\
    \bottomrule
  \end{tabular}
\end{table}

\subsection{Efficiency against the replaced tokenizer}
BlueMagpie-TTS reuses the acoustic stack of VoxCPM2 (Section~\ref{sec:tts}), so on the text side PangolinTokenizer replaces the VoxCPM2 tokenizer. That tokenizer splits every multi-character Chinese token into single characters~\cite{voxcpm2}, because multi-character Chinese tokens otherwise produce unstable audio in its acoustic model. This keeps the audio correct, but it makes the text side close to one token per Chinese character. We compare the two tokenizers on 1{,}000 Taiwanese-Mandarin colloquial sentences. Table~\ref{tab:voxtok} reports the result. PangolinTokenizer uses 30.0\% fewer tokens in total, 19{,}082 against 27{,}245. It uses 0.611 tokens per character, against 0.873, and packs 1.635 characters into each token, against 1.145. The VoxCPM2 rate stays below one because punctuation, digits, and English spans still merge into multi-character tokens. PangolinTokenizer uses fewer tokens on 960 of the 1{,}000 sentences, the VoxCPM2 tokenizer on 22, with 18 ties. This saving is on the text side. It shortens the sequence that the frontend plans over, and it does not change the acoustic stack or the audio.

\begin{table}[t]
\caption{Token efficiency against the replaced tokenizer, on 1{,}000 Taiwanese-Mandarin colloquial sentences. The VoxCPM2 tokenizer is character-level on Chinese by design. Lower tokens per character and higher characters per token are better.}
\label{tab:voxtok}
\centering
\setlength{\tabcolsep}{5pt}
\renewcommand{\arraystretch}{1.15}
\begin{tabular}{@{}lcc@{}}
\toprule
\textbf{Metric} & \textbf{VoxCPM2 tok.} & \textbf{PangolinTok.} \\
\midrule
Total tokens       & 27{,}245 & \textbf{19{,}082} \\
Tokens / sentence  & 27.25 & \textbf{19.08} \\
Tokens / character & 0.873 & \textbf{0.611} \\
Characters / token & 1.145 & \textbf{1.635} \\
\bottomrule
\end{tabular}
\end{table}

\section{Barbet: a Traditional-Chinese Frontend}
\label{sec:lm}
\subsection{From tokenizer to frontend}
The frontend of the base TTS model is a general multilingual language model~\cite{minicpm}. It is not specialized for Traditional Chinese or Taiwan-style mixed text. Two consequences follow. First, it represents Traditional Chinese less efficiently, so its plan for Traditional-Chinese text is weaker. Second, it has seen little intra-sentential Chinese--English mixing, so it plans the switch points poorly. Work on Taiwan-localized language models shows that targeted Traditional-Chinese training raises understanding by a wide margin~\cite{taiwanllm,tmlu}. We therefore replace this frontend with a model centered on Traditional Chinese.

We tie the new frontend to the tokenizer. PangolinTokenizer defines a new vocabulary and a new token distribution, so the base frontend cannot be reused with it. We train a new frontend, Barbet, on PangolinTokenizer. The token savings then propagate to the frontend: a lower token rate gives shorter sequences for the same text, more raw text under a fixed token budget, and, with tied input and output embeddings on a small vocabulary, a smaller parameter share spent on the vocabulary.

Barbet is a research base model, not an instruction-tuned assistant, which is the right form for the frontend role. An instruction-tuned model is shaped to answer prompts, which is the wrong objective for a component that should predict how a sentence is spoken. A base model trained for representation quality keeps the distribution over plausible continuations, which is what the acoustic stack consumes.

\subsection{Architecture and training}
Table~\ref{tab:barbet} lists the configuration. Barbet is a billion-parameter decoder-only model. It is a hybrid decoder that interleaves global attention, sliding-window attention, and a Mamba state-space sequence mixer~\cite{mamba} in a four-layer cycle. It uses grouped-query attention~\cite{gqa} and rotary position embeddings~\cite{rope}, and ties the input and output embeddings. The tokenizer is PangolinTokenizer, padded to 114{,}944 entries for training. Barbet is trained in three stages, with a total budget of about 150 billion tokens. The general stage uses a broad multilingual and code mixture, so the model first learns general language structure. The mid-training stage raises the share of Traditional-Chinese and Taiwan-context text under stricter deduplication and quality filtering, so the model sees cleaner local data where the distribution is shifted. The long-context stage extends the sequence length from 8K to a 262{,}144-token native context in steps, each from the previous checkpoint, so the model learns to use the extended window.

The hybrid design balances quality and cost over long inputs. Global attention captures whole-sequence dependencies. Sliding-window attention covers local context at a lower cost. The Mamba state-space mixer~\cite{mamba} handles long-range structure in linear time, which keeps the cost manageable as the context grows toward 256K tokens. Grouped-query attention~\cite{gqa} shrinks the key-value cache, and rotary position embeddings~\cite{rope} support the long-context extension.

\begin{table}[t]
\caption{The Traditional-Chinese frontend language model (Barbet). It is a hybrid decoder that interleaves attention with a state-space sequence mixer.}
\label{tab:barbet}
\centering
\setlength{\tabcolsep}{5pt}
\renewcommand{\arraystretch}{1.2}
\begin{tabular}{@{}l p{0.56\columnwidth}@{}}
\toprule
\textbf{Property} & \textbf{Value} \\
\midrule
Parameters & $1.09$\,B \\
Decoder layers & 28 \\
Hidden size & 1536 \\
Attention & grouped-query, 16 query / 2 key/value heads \\
Layer pattern & global, sliding-window, Mamba: 4-layer cycle \\
Sliding window & 8192 tokens \\
Position encoding & RoPE, base $10^{7}$ \\
Embeddings & tied input and output \\
Tokenizer & PangolinTokenizer, vocab 114{,}944 (padded) \\
Native context & 262{,}144 tokens (256K) \\
Precision & bf16 \\
\bottomrule
\end{tabular}
\end{table}

\subsection{Standalone quality}
Table~\ref{tab:taide} reports the frontend's standalone quality on TAIDE-14~\cite{taide14}, an open evaluation set from Taiwan's TAIDE project with 14 Traditional-Chinese generation tasks and 140 prompts, as bits per byte. Bits per byte normalizes the loss by the UTF-8 byte count, so it does not depend on the tokenizer vocabulary, which makes the comparison fair across tokenizers. Barbet reaches 0.7488 bits per byte, the lowest among the four compared models, below MiniCPM5-1B at 0.7577 and LLaMA-3.2-1B at 0.9190, and it ranks first on 10 of the 14 tasks. The same run shows the token efficiency of PangolinTokenizer: Barbet encodes the text at 0.2164 tokens per byte, the lowest of the four models, so each token carries more content. A stronger frontend gives the acoustic stack a better plan, which is the part that the next sections build on.

\begin{table}[t]
\caption{Frontend language-model quality on TAIDE-14 (14 tasks, 140 samples), as bits per byte (lower is better). Bits per byte normalizes by UTF-8 bytes, so it is comparable across tokenizers. LFM2.5 is instruction-tuned, so its higher value partly reflects alignment rather than base modeling.}
\label{tab:taide}
\centering
\setlength{\tabcolsep}{5pt}
\renewcommand{\arraystretch}{1.15}
\begin{tabular}{@{}lcc@{}}
\toprule
\textbf{Model} & \textbf{bits/byte} & \textbf{tokens/byte} \\
\midrule
Barbet (ours)        & \textbf{0.7488} & \textbf{0.2164} \\
MiniCPM5-1B          & 0.7577 & 0.2804 \\
LLaMA-3.2-1B         & 0.9190 & 0.2960 \\
LFM2.5-1.2B (inst.)  & 2.0082 & 0.3408 \\
\bottomrule
\end{tabular}
\end{table}

\section{BlueMagpie-TTS}
\label{sec:tts}
\subsection{Bridge module and two-stage training}
BlueMagpie-TTS attaches Barbet to the reused VoxCPM2 acoustic stack. The two parts were pretrained separately, so their representations do not match: Barbet emits 1536-dimensional hidden states in its own space, and the acoustic stack expects inputs in the space of its original text-semantic frontend. A bridge module resolves this mismatch. It projects the hidden states of Barbet into the input space of the acoustic stack, so the two pretrained parts can exchange information. The bridge is an RMSNorm and a linear projection, followed by one residual SwiGLU block whose output projection is zero-initialized. At initialization the bridge reduces to the normalized linear map, a stable warm start. The bridge is the only component learned from scratch, which is what makes the recipe cheap to apply to a new target.

Training proceeds in two stages, and the order protects the pretrained weights. In the first stage, we freeze Barbet and the acoustic stack and train only the bridge by hidden-space distillation: its output is matched to the hidden states that the original VoxCPM2 frontend produces for the same text. Training the bridge alone protects both pretrained parts from the noisy gradients of a fresh module. Once the bridge aligns the two spaces, the second stage unfreezes the assembled model and jointly fine-tunes Barbet, the bridge, and the acoustic stack on target-voice speech, so the plan and the rendering adapt to each other. Because every part except the bridge starts from pretrained weights, this fine-tuning is short relative to training a TTS model from scratch.

\subsection{Inference and control}
The model exposes one generation interface with several modes: plain synthesis; voice cloning from a reference clip; speaker control from a precomputed speaker vector; and streaming output. Two parameters matter most, shown in Table~\ref{tab:params}. The guidance strength controls how closely the output follows the text. The number of diffusion steps trades quality for speed. We tune these two on the Taiwan-localized test set, keep every other setting at its default, and apply the same values to every system in Section~\ref{sec:exp}, so the tuning does not favor one system. A higher guidance strength helps the model commit to the English tokens at switch points, so we raise it from 2.0 to 2.8. The output is 48~kHz mono.

\begin{table}[t]
\caption{Key generation parameters. We tune two, keep the rest at their defaults, and use the same values for every system.}
\label{tab:params}
\centering
\setlength{\tabcolsep}{5pt}
\renewcommand{\arraystretch}{1.15}
\begin{tabular}{lccp{0.27\columnwidth}}
\toprule
\textbf{Parameter} & \textbf{Default} & \textbf{Used} & \textbf{Effect} \\
\midrule
Guidance strength & 2.0 & \textbf{2.8} & higher follows the text more closely \\
Diffusion steps & 10 & \textbf{9} & more steps give quality at lower speed \\
Bad-case retry & off & \textbf{on} & re-samples a detected anomalous output \\
Sample rate & \multicolumn{2}{c}{48\,kHz mono} & fixed output format \\
\bottomrule
\end{tabular}
\end{table}

\subsection{Responsible use}
Voice cloning and speaker control can imitate a target voice, so they need consent. We use voices only with permission. The default voice comes from one of the authors, recorded with consent. We do not provide a cloning interface for non-consenting targets, and deployments should keep this constraint.

\section{Experiments}
\label{sec:exp}
\subsection{Evaluation protocol}
Listening tests are costly and hard to reproduce. We measure intelligibility with a synthesize-then-recognize loop. The loop runs in one direction. BlueMagpie-TTS synthesizes the input text to speech. The Taiwanese-Mandarin ASR model Breeze-ASR-25~\cite{breezeasr} transcribes the speech back to a hypothesis. We then compare the hypothesis with the input text. We report CER, the character-level edit rate between the input and the transcript. We use CER rather than WER, because Mandarin has no word boundaries, so a word-level metric would depend on a segmenter and inherit its errors. Character-level scoring is standard for Mandarin speech, and it applies directly to the English spans of a code-switching sentence. Normalization lowercases English, removes punctuation, and maps full-width forms to half-width, so that scoring reflects pronunciation rather than orthography. The ASR makes its own errors, so the metric reflects intelligibility relative to this ASR, but we use the same ASR for every system, so the comparison stays fair. We choose this ASR because it targets Taiwanese Mandarin and code-switching, so it is a strict judge for the phenomena we test.

The loop is a proxy. It measures whether the spoken output is recoverable as the intended text, which is intelligibility. It does not measure naturalness or speaker similarity directly, which is why we complement it with the listening study in Section~\ref{sec:pref}. A high error rate signals a real problem, a dropped or blurred span, while a low error rate is necessary but not sufficient for good speech.

\subsection{Test set}
We use a set of 1{,}000 Taiwan-localized sentences. Each sentence reflects common Taiwanese Mandarin usage. Many contain English words, abbreviations, numbers, or proper nouns, the cases where Taiwanese speech applications tend to fail. The English spans fall into four types: full words, for example Transformer; abbreviations and acronyms, for example LLM and API; numbers and units; and proper nouns, for example OpenAI. Each sentence is short, one or two clauses, so when a switch point is present it dominates the example, and the metric is sensitive to it. This is an internal test set, not a public benchmark, so the numbers support relative comparison. Table~\ref{tab:examples} shows representative sentences, including a failure case. Three systems appear in the comparison. The zero-shot base is the unmodified VoxCPM2 given a reference speech, with its original text-semantic frontend. The fine-tuned base starts from the same system and fine-tunes the acoustic stack on the target-voice data. BlueMagpie-TTS replaces the frontend with Barbet through the bridge and applies the same target-voice fine-tuning. We decode all systems with the generation parameters in Table~\ref{tab:params}, on the same hardware. We fix the random seed where the interface allows it, and the bad-case retry is the only source of run-to-run variation.

\begin{table}[t]
\caption{Representative test sentences. Mandarin matrix with embedded English tokens (bold). The last row is a failure case.}
\label{tab:examples}
\centering
\renewcommand{\arraystretch}{1.25}
\begin{tabular}{p{0.94\columnwidth}}
\toprule
\zh{這個} \textbf{Transformer} \zh{架構，其實就是現在所有聊天機器人的底層。}\\
\textbf{OpenAI} \zh{我都直接念英文，可是} \textbf{TTS} \zh{常把} open \zh{跟} A I \zh{黏在一起變成怪音。}\\
\midrule
\zh{我把整篇逐字稿丟給} \textbf{LLM}\zh{，叫它幫我整理成三個重點。}\\
\footnotesize\itshape failure: the token ``LLM'' is articulated unclearly, and the ASR transcribes it as ``LOL and''.\\
\bottomrule
\end{tabular}
\end{table}

\subsection{Main results}
\begin{figure}[t]
\centering
\includegraphics[width=\columnwidth]{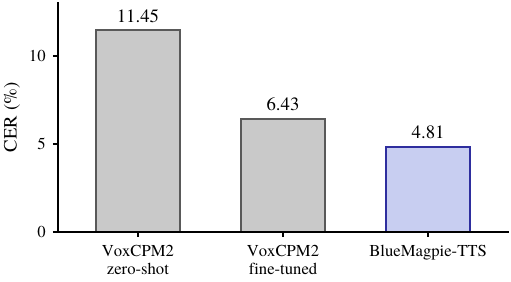}
\caption{CER on the 1{,}000-sentence Taiwan-localized test set. Fine-tuning the VoxCPM2 acoustic stack to the target voice lowers CER from 11.45\% to 6.43\%. Replacing its frontend with Barbet lowers CER further to 4.81\%.}
\label{fig:bars}
\end{figure}

Table~\ref{tab:main} reports the main result. The zero-shot base, VoxCPM2 with its original frontend and a reference speech, reaches 11.45\% CER. BlueMagpie-TTS reaches 4.81\% CER, a relative reduction of 58.0\%. Figure~\ref{fig:bars} visualizes the three systems.

A larger gap does not by itself isolate the frontend, because BlueMagpie-TTS also fine-tunes its acoustic stack to the target voice. We add a fine-tuned baseline to separate the two effects: VoxCPM2 with its original frontend, fine-tuned on the same target-voice data with the same hyperparameters. This baseline reaches 6.43\% CER. Fine-tuning the acoustic stack recovers part of the gap, from 11.45\% to 6.43\%, and BlueMagpie-TTS reaches 4.81\%, a further 25.2\% reduction. The two systems share the same acoustic stack and the same target-voice fine-tuning, and differ only in the frontend and its bridge, so this remaining gain comes from the Taiwan-localized text side, not from voice adaptation.

By inspection, the base tends to drop or blur the English span, which the ASR then mis-recognizes, while BlueMagpie-TTS renders the span as a clear English pronunciation inside the Mandarin matrix. The numeric gap matches what we hear on the switch-heavy sentences.

The test set is built from locally relevant cases on purpose, so it tests the accent and the switch points rather than providing an average-case measure of the system.

\begin{table}[t]
\caption{Intelligibility on the 1{,}000-sentence Taiwan-localized test set, by the synthesize-then-recognize loop. Lower is better. CER is the primary metric; the fine-tuned baseline is scored on CER. All systems use the VoxCPM2 acoustic stack and the same ASR.}
\label{tab:main}
\centering
\setlength{\tabcolsep}{5pt}
\begin{tabular}{@{}p{0.66\columnwidth}c@{}}
\toprule
\textbf{System} & \textbf{CER (\%)} \\
\midrule
Base: VoxCPM2, zero-shot (reference) & 11.45 \\
    Base: VoxCPM2, fine-tuned to voice & 6.43 \\
\textbf{BlueMagpie-TTS} & \textbf{4.81} \\
\midrule
Reduction vs.\ zero-shot & 58.0 \\
    Reduction vs.\ fine-tuned & 25.2 \\
\bottomrule
\end{tabular}
\end{table}

\subsection{Human preference}
\label{sec:pref}
The synthesize-then-recognize loop measures intelligibility, not naturalness. We add a listening study that compares BlueMagpie-TTS and the zero-shot base directly. For each sentence, we play the two outputs side by side in random order, and a listener picks the one that sounds more natural for Taiwanese Mandarin, or marks a tie. We collect three ratings per sentence on 500 randomly sampled sentences from the test set, from ten native speakers of Taiwanese Mandarin, and take the majority vote. Table~\ref{tab:pref} reports the result. Listeners prefer BlueMagpie-TTS on 65.6\% of the sentences, against 17.0\% for VoxCPM2, with the rest showing no clear preference. Among the 413 decisive sentences, BlueMagpie-TTS wins 79.4\%, and a two-sided sign test rejects equal preference ($p<10^{-30}$). The vote level shows the same pattern: of all 1{,}500 votes, 64.8\% favor BlueMagpie-TTS, 20.4\% favor VoxCPM2, and 14.8\% are ties. The study uses the same sentences as the intelligibility study, and the two measures agree: the frontend swap that lowers CER also wins the direct comparison.

\begin{table}[t]
\caption{Human preference on 500 sentences from the 1{,}000-sentence test set, three listeners per sentence, by majority vote. ``No clear preference'' combines tie majorities (44) and split votes (43).}
\label{tab:pref}
\centering
\setlength{\tabcolsep}{6pt}
\renewcommand{\arraystretch}{1.15}
\begin{tabular}{@{}lcc@{}}
\toprule
\textbf{Outcome} & \textbf{Sentences} & \textbf{Share} \\
\midrule
\textbf{BlueMagpie-TTS preferred} & \textbf{328} & \textbf{65.6\%} \\
VoxCPM2 preferred & 85 & 17.0\% \\
No clear preference & 87 & 17.4\% \\
\bottomrule
\end{tabular}
\end{table}

\subsection{Speed and error analysis}
We measure the inverse real-time factor (xRT), the seconds of audio produced per second of compute. On our hardware, the median xRT is 4.75 and the maximum is 5.29, so the model synthesizes about five times faster than real time, which supports offline synthesis and real-time streaming playback. We keep nine diffusion steps, one below the default, because the extra step gives little quality at a measurable speed cost on this test set. The main remaining weakness is short English abbreviations inside a Mandarin matrix. The token ``LLM'' is not articulated cleanly, and the ASR transcribes it as ``LOL and'' (Table~\ref{tab:examples}). Single letters and short acronyms, spelled out letter by letter, are the hardest, because the boundary is short and the acronym carries little acoustic context. Longer English words and frequent proper nouns, such as ``OpenAI'' and ``Transformer'', are rendered and recovered cleanly. Numbers and units are an intermediate case. This pattern points to a targeted fix: more acronym-rich code-switching data for the frontend, rather than any change to the acoustic stack.

\subsection{Efficiency of the text side}
PangolinTokenizer's token efficiency also reaches the TTS: for the same input, Barbet plans over a token sequence about 8\% to 13\% shorter than the same model on a similar-size multilingual tokenizer (Table~\ref{tab:compression}). A shorter plan lowers inference cost and KV-cache pressure. The smaller vocabulary, 114{,}944 against 200K to 318K, also reduces the embedding and output-projection parameters under tied embeddings.

\section{Discussion}
\label{sec:disc}
The stack keeps one property from end to end: the text side is token-efficient at the tokenizer, the language model, and the TTS. PangolinTokenizer lowers the token cost at a small vocabulary, Barbet inherits this efficiency, and BlueMagpie-TTS changes only the frontend of a strong acoustic stack. The comparison in Table~\ref{tab:main} is controlled: the fine-tuned base and BlueMagpie-TTS differ only in the text frontend and its bridge, so the remaining gap is a frontend effect: a substantial part of accent and code-switching quality is set by the frontend, not by acoustic rendering. Prior code-switching TTS often retrains acoustic models for the target language pair~\cite{cstts}; our result suggests that the switch-point quality is mostly set by the text frontend when the acoustic stack is strong.

Training an acoustic stack from scratch needs large-scale paired speech. Our recipe reuses that stack: the tokenizer and the frontend train on cheaper text, and the bridge plus joint fine-tuning need only a short target-voice run.

Two limitations point to future work: the design couples the tokenizer and the frontend, so their contributions to CER are not separated; and the short test sentences do not exercise Barbet's 262K context, leaving long-form synthesis open.

\section{Conclusion}
\label{sec:concl}
We presented a Taiwan-localized speech stack built around token efficiency: PangolinTokenizer reaches the lowest token rate (0.485 tokens/character) and the highest efficiency among eight tokenizers. Barbet, trained on it, is the frontend and ranks first among the compared models on TAIDE-14. BlueMagpie-TTS attaches Barbet to the reused VoxCPM2 acoustic stack through a bridge, lowers CER from 11.45\% to 4.81\% on a 1{,}000-sentence Taiwan-localized test set, stays 25.2\% below a voice-fine-tuned baseline, and listeners prefer it on 65.6\% of sentences.

\balance
\bibliographystyle{IEEEtran}
\bibliography{merged}

\end{document}